\documentclass{PoS}

\usepackage[utf8]{inputenc}
\usepackage{amsmath,amsfonts,amssymb,mathrsfs,fixmath}

\usepackage[export]{adjustbox} %http://ctan.org/pkg/adjustbox

\usepackage{etoolbox}
\newtoggle{arxiv}
\toggletrue{arxiv}

\newcommand{\mycite}[2]{\iftoggle{arxiv}{\protect\cite{#1,#2}}{\protect\cite{#1}}}
\newcommand{\ifarxiv}[1]{\iftoggle{arxiv}{#1}{}}

\newcommand{\comment}[1]{}
\newcommand{\lr}[1]{ \left( #1 \right) }
\newcommand{\lrs}[1]{ \left[ #1 \right] }

\newcommand{\vev}[1]{ \langle \, #1 \, \rangle }

\newcommand{\tr}{ {\rm Tr} \, }

\newcommand{\const}{ {\rm const}}
\renewcommand{\det}[1]{ {\rm det} \left( #1 \right) }

\newcommand{\expa}[1]{ \exp{\left( #1 \right)} }
\newcommand{\diag}[1]{ {\rm diag} \, \left( #1 \right) }

\title{Feasibility of Diagrammatic Monte-Carlo based on weak-coupling expansion in asymptotically free theories: case study of $O\lr{N}$ sigma-model in the large-$N$ limit}

\ShortTitle{DiagMC for large-$N$ $O\lr{N}$ sigma-model}

\author{\speaker{P.~V.~Buividovich}\thanks{This work was supported by the S.~Kowalevskaja award from the Alexander von Humboldt foundation. The author is grateful to G.~Dunne, T.~Sulejmanpasic and M.~Unsal for interesting and stimulating discussions.}\\
        Regensburg University, D-93053 Regensburg, Germany\\
        E-mail: \email{pavel.buividovich@physik.uni-regensburg.de}}

\abstract{We discuss the feasibility of applying Diagrammatic Monte-Carlo algorithms to the weak-coupling expansions of asymptotically free quantum field theories, taking the large-$N$ limit of the $O\lr{N}$ sigma-model as the simplest example where exact results are available. We use stereographic mapping from the sphere to the real plane to set up the perturbation theory, which results in a small bare mass term proportional to the coupling $\lambda$. Counting the powers of coupling associated with higher-order interaction vertices, we arrive at the double-series representation for the dynamically generated mass gap in powers of both $\lambda$ and $\log(\lambda)$, which converges quite quickly to the exact non-perturbative answer. We also demonstrate that it is feasible to obtain the coefficients of these double series by a Monte-Carlo sampling in the space of Feynman diagrams. In particular, the sign problem of such sampling becomes milder at small $\lambda$, that is, close to the continuum limit.}

\FullConference{The 33rd International Symposium on Lattice Field Theory\\
		14 -18 July 2015\\
		Kobe International Conference Center, Kobe, Japan*}

\begin{document}
\sloppy

\section{Introduction}
\label{sec:intro}

 Diagrammatic Monte-Carlo (DiagMC) algorithms \mycite{Prokofev:08:1}{Prokofev:98:1} which stochastically sample strong- or weak-coupling expansion diagrams provide a useful alternative to the standard Monte-Carlo algorithms which are based on stochastic sampling of field configurations. In recent years, DiagMC algorithms attracted a lot of interest from lattice QCD community as a prospective tool for reducing the sign problem in lattice QCD simulations at finite chemical potential \mycite{DeForcrand:10:1,Philipsen:13:1}{deForcrand:14:1,Rossi:84:1,Wolff:85:1}. So far, all attempts to apply DiagMC to non-Abelian lattice gauge theories are based on the few lowest orders of the strong-coupling expansion of the QCD partition function. This approach turns out to be very efficient, since already the lowest order of strong-coupling expansion captures quark confinement, a fundamental feature of QCD. However, these algorithms become in general inapplicable as one approaches the continuum limit, since one has to take into account more and more terms in the strong coupling expansion. \ifarxiv{For Abelian gauge theories \cite{Wolff:10:3,Gattringer:15:1,Gattringer:13:1} as well as for $O\lr{N}$ and $CP^N$ sigma-models \cite{Wolff:09:1,Wolff:10:1,Sulejmanpasic:15:1}, strong-coupling expansion can be organized into series with positive coefficients, which can be sampled to arbitrary order using the worm-type DiagMC algorithms. For finite-volume systems, such sampling yields accurate results even close to the continuum limit (or the quantum phase transition).}

 Unfortunately, up to now no efficient ways of automated and systematically improvable stochastic sampling of strong-coupling expansions in lattice systems with $SU\lr{N}$-valued degrees of freedom (including lattice QCD) are known. Moreover, one can expect that at large orders of strong-coupling expansion some terms in the series become negative, thus leading to a (real) sign problem in DiagMC. This sign problem should become quite severe close to the continuum, where large factors proportional to negative powers of coupling should cancel to yield a small result close to unity (e.g. for the mean plaquette). \ifarxiv{The situation might become even more problematic in the large-$N$ limit, where the strong- and weak-coupling regimes are expected to be separated by a quantum phase transition \cite{Gross:80:1}.}

 In this situation it seems tempting to devise DiagMC algorithms which are based on the conventional weak-coupling perturbation theory, where diagrammatic rules are comparatively easy to obtain. Before turning to real simulations of lattice field theories with non-Abelian degrees of freedom \ifarxiv{(such as principal chiral models or non-Abelian lattice gauge theories)}, in these Proceedings we discuss the feasibility of such approach on the simplest example of the exactly solvable $O\lr{N}$ sigma-model on the lattice in the large-$N$ limit. \ifarxiv{This model has been traditionally considered as the ``minimal working example'' for nontrivial features of the perturbative expansions in more complicated asymptotically free QFTs \cite{David:81:1,David:82:1,Novikov:84:1,Braun:98:1}.} At the leading order in $1/N$ expansion, only Feynman diagrams of ``cactus'' topology \ifarxiv{(see Fig.~\ref{fig:skl_convergence} for an illustration)} contribute to the weak-coupling expansion of this model, which allows one to obtain high-order expansion coefficients by a simple recursive procedure. Our aim here is to study whether these coefficients could be in principle obtained by Monte-Carlo sampling in the space of \ifarxiv{``cactus''-like }Feynman diagrams.\ifarxiv{ The space of such diagrams is certainly a subspace of a larger space of diagrams which should be sampled in generic DiagMC simulations of e.g. principal chiral models or non-Abelian lattice gauge theories.}

 Having started with such a motivation, we immediately face the following conceptual problem: in asymptotically free QFTs the mass gap is typically non-perturbative and has the form $m^2 \sim e^{-\beta_0/\lambda}$, where $\beta_0$ is the zeroth order term in the expansion of the beta-function and $\lambda$ is the t'Hooft coupling constant. This statement is true also in the large-$N$ limit, where the number of Feynman diagrams which contribute at a given order of $1/N$ expansion is known to grow exponentially with diagram order. Now if at the leading order of the $1/N$ expansion the contribution of all diagrams is finite\ifarxiv{\footnote{Since we work on the lattice, UV divergences are all regulated}}, the result of summation over them should be analytic in $\lambda$ at least in some vicinity of $\lambda = 0$. Thus the only way in which the non-perturbative scale can emerge is through IR divergences in some of the diagrams. However, DiagMC simulations would be problematic if not impossible with such divergences. \ifarxiv{One possible way to deal with them would be to introduce some IR regulator and extrapolate it to zero after the simulations, which introduces yet another potential source of systematic errors. Such approach is used, e.g., in numerical stochastic perturbation theory \cite{DiRenzo:04:1} and allows one to observe the IR renormalon non-summability of the perturbative expansion \cite{Pineda:14:1,Pineda:15:1}, which should persist even in the large-$N$ limit. Another possible way is to use some sort of ``bold'' DiagMC algorithm \cite{Pollet:10:1,Davody:13:1} where the diagrams are partially re-summed so that the bare mass term appears. So far, however, ``bold'' DiagMC algorithms typically rely on some truncation of vertices in the Schwinger-Dyson equations  and thus appear to be quite model-dependent.}

 In these Proceedings we show that certain parameterizations of field variables remove IR divergences in a self-consistent way even from ``undressed'' Feynman diagrams by introducing a small bare mass proportional to the coupling $\lambda$. The resulting series, however, no longer have a conventional form of power series in $\lambda$, but are rather double series both in $\lambda$ and in $\log{\lambda}$\ifarxiv{, somewhat reminiscent of trans-series}. Here we take a closer look at the structure of such series and demonstrate numerically that they converge to the exact nonperturbative answer. We also show that the coefficients of these series can be obtained by a Monte-Carlo sampling in the space of Feynman diagrams, with the sign problem becoming milder in the continuum limit.

\section{Large-$N$ $O\lr{N}$ sigma-model in stereographic coordinates}
\label{sec:stereographic_coords_def}

 Field variables in the $O\lr{N}$ sigma model are $N$-component unit vectors $n_{a \, x}$, $a = 0 \ldots N-1$ attached to the sites of the two-dimensional square lattice, which we label by $x$. The Euclidean partition function is given by
\begin{eqnarray}
\label{on_partition1}
 \mathcal{Z} = \int\mathcal{D}n_x \expa{-\frac{N}{2 \lambda} \sum\limits_{x,y,a} D_{xy} n_{a \, x} \, n_{a \, y} } ,
\end{eqnarray}
where $D_{x,y} = 4 \delta_{x,y} - \sum\limits_{\mu=1,2} \delta_{x, y + \hat{\mu}} - \sum\limits_{\mu=1,2} \delta_{x, y - \hat{\mu}}$ is the lattice Laplacian. Superficially, in the weak-coupling limit $\lambda \rightarrow 0$ all the vectors $n_x$ should align in one direction, thus spontaneously breaking the global $O\lr{N}$ symmetry of the model and leaving $N-1$ massless Goldstone modes in the spectrum. However, by virtue of the Mermin-Wagner theorem such spontaneous symmetry breaking cannot occur in two dimensions, and the $O\lr{N}$ symmetry remains unbroken for all values of $\lambda$. \ifarxiv{Instead of $N-1$ massless Goldstones, the model describes $N$ massive free particles with the mass $m^2 \equiv \sigma \sim e^{-\frac{4 \pi}{\lambda}}$ which is manifestly non-perturbative in $\lambda$.}

 The standard way to arrive at the weak-coupling perturbative expansion is, however, to expand around the minimum-energy configuration with all $n_x$ aligned in one direction, say, in the direction with $a = 0$.\ifarxiv{ A consistent perturbative expansion should then restore the $O\lr{N}$ symmetry.} In order to perform such expansion, we need to introduce some coordinates $\phi_{i \, x}$, $i = 1 \ldots N-1$ parameterizing the vectors $n_x$, such that $\phi_{i \, x} = 0$ corresponds to the minimum-energy configuration with $n_x = \const$. In this work, we use the stereographic mapping
\begin{eqnarray}
\label{stereographic_mapping_perturbative}
 n_{0 \, x} = \frac{1 - \frac{\lambda}{4} \phi_x^2}{1 + \frac{\lambda}{4} \phi_x^2},
 \quad
 n_{i \, x} = \frac{\sqrt{\lambda} \, \phi_{i \, x}}{1 + \frac{\lambda}{4} \phi_x^2},
 \quad
 \phi_x^2 \equiv \sum\limits_{i} \phi_{i \, x} \phi_{i \, x}
\end{eqnarray}
from the whole real space $\mathbb{R}^{N-1}$ to the whole sphere $S_N$ \ifarxiv{\footnote{More precisely, the point with $n_{0} = -1$ should be excluded from the mapping for topological reasons, but this point has measure zero in the path integral and is therefore irrelevant.}}. The integration measure on $S_N$ in terms of stereographic coordinates $\phi$ reads \ifarxiv{(see Appendix \ref{apdx:sn_stereo_mapping_measure} for the derivation)}
\begin{eqnarray}
\label{SN_integration_measure_stereo}
 \mathcal{D}n_x = \mathcal{D}\phi_x \lr{1 + \frac{\lambda}{4} \phi_x^2}^{-N} .
\end{eqnarray}
Using (\ref{stereographic_mapping_perturbative}) and (\ref{SN_integration_measure_stereo}), we can express the partition function as an integral over the fields $\phi_x$. In order to carry out the perturbative expansion, it is convenient to explicitly separate the ``free'' part of the action which is quadratic in $\phi_x$ from the interaction part, which contains higher powers of $\phi_x$ multiplied by some positive powers of the coupling constant $\lambda$:
\begin{eqnarray}
\label{on_partition_perturbative}
 \mathcal{Z} = \int\mathcal{D}\phi_x
 \expa{-N \log\lr{1 + \frac{\lambda}{4} \phi_x^2}
 - \frac{N}{2 \lambda}
   \sum\limits_{x, y} D_{x y} \frac{\lr{1 - \frac{\lambda}{4} \phi_x^2 } \lr{1 - \frac{\lambda}{4} \phi_y^2 } + \lambda \lr{\phi_x \cdot \phi_y} }{\lr{1 + \frac{\lambda}{4} \phi_x^2 } \lr{1 + \frac{\lambda}{4} \phi_y^2 }}
 }
 = \nonumber \\  =
 \int\mathcal{D}\phi_x \expa{-\frac{1}{2} \sum\limits_{x,y} \lr{D_{x y} + \frac{\lambda}{2} \delta_{x y} } {\phi_x \cdot \phi_y} \, + \, S_I\lrs{\phi}} , \hspace{1cm}
 \nonumber \\
 S_I\lrs{\phi} =
 \sum\limits_{\substack{k,l=0\\k+l\neq 0}}^{+\infty} \frac{\lr{-1}^{k+l} \lambda^{k+l}}{2 \cdot 4^{k+l}}
   \sum\limits_{x, y} D_{x y} \lr{\phi_x^2}^k \lr{\phi_y^2}^l \lr{\phi_x \cdot \phi_y}
 +
\sum\limits_{k=2}^{+\infty} \frac{\lr{-1}^{k-1} \lambda^k}{4^k \, k} \sum\limits_x \lr{\phi_x^2}^k
 , \hspace{1cm}
\end{eqnarray}
where $\phi_x \cdot \phi_y \equiv \sum\limits_i \phi_{i\,x} \cdot \phi_{i\,y}$ and in the last expression for $S_I\lrs{\phi}$ we have dropped the summands of the form $\sum\limits_{x,y} D_{xy} \lr{\phi_x^2}^k \lr{\phi_y^2}^l$ which are effectively zero in the large-$N$ limit by virtue of factorization and translational invariance. We thus see that due to the nontrivial integration measure in terms of the fields $\phi_x$ the bare ``mass term'' $\lambda/2$ appears in the quadratic part of the action. \ifarxiv{The appearance of such a bare mass term is quite generic and perhaps the only mapping from $\mathbb{R}^{N-1}$ to $S_N$ in which it does not appear is $n_0 = \sqrt{1 - c \phi^2}$, $n_i = \sqrt{c} \phi_i$, with $c$ being some constant \cite{David:81:1}.}

 Having written the partition function and the action in the form (\ref{on_partition_perturbative}), now we can regard our $O\lr{N}$ sigma-model simply as a large-$N$ quantum scalar field theory with infinitely many interaction vertices. The standard way to the exact solution of such theories is to consider the Schwinger-Dyson equations, which in the large-$N$ limit reduce to a single equation on the two-point correlation function $\vev{\phi_x \cdot \phi_y}$\ifarxiv{ (for completeness, we derive this equation in Appendix \ref{apdx:sd_equations})}. All correlators with larger number of field operators reduce to the products of $\vev{\phi_x \cdot \phi_y}$. Using the momentum-space representation of the two-point function $\vev{\phi_x \cdot \phi_y} = \int \frac{d^2 p}{\lr{2 \pi}^2} \, G\lr{p} \, e^{i p \lr{x - y}}$, where integration goes over the square Brillouin zone $p_1, p_2 \in \lrs{-\pi, \pi}$, we can write the Schwinger-Dyson equations as
\begin{eqnarray}
\label{perturbative_G_stereo_msummed}
 G^{-1}\lr{p} = D\lr{p} + \sigma_{0}
 - %\nonumber \\ -
 D\lr{p} \frac{\xi \lr{2 + \xi}}{\lr{1 + \xi}^2}
 -
 \frac{\lambda}{2} \frac{\xi}{1 + \xi}
 - %\nonumber \\ -
 \frac{\lambda}{2} \frac{1}{\lr{1 + \xi}^3} \lrs{D G}_{xx} ,
\end{eqnarray}
where $D\lr{p} = 4 \sin^{2}\lr{p_1/2} + 4 \sin^{2}\lr{p_2/2}$ is the momentum-space representation of the lattice Laplacian in (\ref{on_partition1}) and we have denoted $\sigma_0 \equiv \lambda/2$ and $\xi \equiv \frac{\lambda}{4} \, \vev{\phi_x^2} = \frac{\lambda}{4} \, \int\frac{d^2 p}{\lr{2 \pi}^2} \, G\lr{p}$ and $\lrs{D G}_{xx} = \int\frac{d^2 p}{\lr{2 \pi}^2} \, D\lr{p} \, G\lr{p}$. \ifarxiv{By virtue of translational invariance, these quantities do not depend on $x$. The first two summands on the r.h.s. of this equation correspond to the quadratic part of the action in (\ref{on_partition_perturbative}).}
The notation $\sigma_0 \equiv \lambda/2$ \ifarxiv{for the bare self-energy term} is introduced to facilitate the formal counting of the positive powers of $\lambda$ associated with the vertices (rather than lines) of the Feynman diagrams, which we perform in the next Section.

 From the form of the equation (\ref{perturbative_G_stereo_msummed}) one can immediately conclude that the momentum-space two-point function $G\lr{p}$ should have the form of the free scalar field propagator with the wave-function renormalization factor $z^2\lr{\lambda}$ and the self-energy $\sigma\lr{\lambda}$:
\iftoggle{arxiv}{\begin{eqnarray}}{$}
 G\lr{p; \lambda} = \frac{z^2\lr{\lambda}}{D\lr{p} + \sigma\lr{\lambda}} .
\iftoggle{arxiv}{\end{eqnarray}}{$}
\ifarxiv{In other words, the large-$N$ limit of our model is completely characterized by the two functions $z\lr{\lambda}$ and $\sigma\lr{\lambda}$. In order to shorten the notation, in what follows we omit the arguments of $\sigma$ and $z$.} From the Schwinger-Dyson equations (\ref{perturbative_G_stereo_msummed}) one can readily deduce the following equations for $z$ and $\sigma$:
\begin{eqnarray}
\label{SD_zsigma_perturbative}
 \sigma = \sigma_0 z^2 - \frac{\lambda}{2} z^2 + \frac{\lambda}{2} z \sigma I_0\lr{\sigma},
 \quad
 z = 1 + \frac{\lambda}{4} z^2 I_0\lr{\sigma} ,
\end{eqnarray}
where we have defined
\iftoggle{arxiv}{\begin{eqnarray}
\label{I0_def}}{$}
 I_0\lr{\sigma} =
 \iftoggle{arxiv}{
  \int\limits_{-\pi}^{\pi} \int\limits_{-\pi}^{\pi} \frac{dp_1 \, dp_2}{\lr{2 \pi}^2} \frac{1}{D\lr{p} + \sigma}
 }{
  \int \frac{d^2 p}{\lr{2 \pi}^2} \frac{1}{D\lr{p} + \sigma}
 }
 = - \frac{1}{4 \pi} \, \log\lr{\frac{\sigma}{32}}\lr{1 + O\lr{\sigma}} .
\iftoggle{arxiv}{\end{eqnarray}}{$}
The last equality holds for sufficiently small values of $\sigma$, and \ifarxiv{since we are mainly interested in the continuum limit where $\sigma$ indeed should be very small, }we will use the latter expression with $O\lr{\sigma}$ terms omitted. Substituting $\sigma_0 = \lambda/2$ into the equations (\ref{SD_zsigma_perturbative}), we obtain the following exact solutions for $z\lr{\lambda}$ and $\sigma\lr{\lambda}$:
\begin{eqnarray}
\label{on_exact_solution}
 z\lr{\lambda} = 2, \quad \lambda I_0\lr{\sigma} = 1 \, \Rightarrow \, \sigma\lr{\lambda} = 32 \expa{-\frac{4 \pi}{\lambda}} ,
\end{eqnarray}
which explicitly demonstrates that the dynamically generated mass term $m\lr{\lambda} = \sqrt{\sigma\lr{\lambda}}$ is non-perturbative in the coupling constant $\lambda$. \ifarxiv{Using this exact solution, it is also easy to check that the $O\lr{N}$ symmetry is indeed restored with our parameterization of $S_N$ in terms of stereographic coordinates (\ref{stereographic_mapping_perturbative}). Namely, the expectation value of $n_{0 \, x}$ is zero, and thus the model forgets about the preferred direction $a = 0$ which we have chosen to set up our perturbative expansion.}

\section{Formal perturbative expansion using stereographic coordinates}
\label{sec:stereographic_expansion}
\ifarxiv{\subsection{Double series in $\lambda$ and $\log\lr{\lambda}$}
\label{subsec:double_series}}

 The standard way to appear at the perturbative expansion is to expand the exponent of the interacting part of the action in power series and to explicitly integrate the resulting polynomials of field variables with the Gaussian weight containing the free part of the action. \ifarxiv{In this process, one can treat the ``bare mass'' term $\frac{\lambda}{2} \phi_x^2$ in two ways. First, if one is interested in the conventional form of perturbative series which contain only positive powers of $\lambda$, one can include this term in the interacting part of the action and leave only the massless kinetic term in the free action. In this case, however, perturbation theory would be plagued by infrared problems and one would need to introduce some some IR regulator in the form of e.g. twisted boundary conditions, as is commonly done in numerical stochastic perturbation theory \cite{DiRenzo:04:1}}.

 In this work, we regard the term $\frac{\lambda}{2} \phi_x^2$ as a part of the free action, so that the free propagator becomes massive and thus IR divergences in \ifarxiv{bare }perturbation theory are cured.\ifarxiv{ In this case, however, the power counting of $\lambda$ is not well defined, since not only the vertices, but also the bare lines of Feynman diagrams have nontrivial dependence on $\lambda$.} In order to avoid any ambiguities in \ifarxiv{such} power counting, let us for the moment forget about the relation between $\sigma_0$ and $\lambda$ in the Schwinger-Dyson equations (\ref{perturbative_G_stereo_msummed}). We are now looking for the solution in form of the formal power series in $\lambda$:
\begin{eqnarray}
\label{formal_power_series}
 z\lr{\lambda}      = \sum\limits_{k=0}^{+\infty} z_k\lr{\sigma_0} \lambda^k, \quad
 \sigma\lr{\lambda} = \sum\limits_{k=0}^{+\infty} \sigma_k\lr{\sigma_0} \lambda^k ,
\end{eqnarray}
with $z_0 = 1$. From the above equations one can recursively express $z_k\lr{\sigma_0}$ and $\sigma_k\lr{\sigma_0}$ in terms of all lower-order coefficients $z_l\lr{\sigma_0}$, $\sigma_l\lr{\sigma_0}$ with $l < k$. Such recursion is mathematically equivalent to the summation of all the Feynman diagrams of a given order which contribute to the two-point function $G\lr{p}$. \ifarxiv{In the large-$N$ limit of our model, only the diagrams with the ``cactus'' topology contribute, see Fig.~\ref{fig:skl_convergence} for an illustration.} In the process of recursion, we still treat $\sigma_0$ as a parameter independent of $\lambda$\ifarxiv{, that is, we do not take into account the equality $\sigma_0 = \lambda/2$ when expanding (\ref{SD_zsigma_perturbative}) in powers of $\lambda$}.

\iftoggle{arxiv}{
\begin{figure*}[h!t]
  \centering
  \newlength{\skpll}\setlength{\skpll}{5.29cm}
  \hspace{1.5cm}
  \includegraphics[width=\skpll,valign=c]{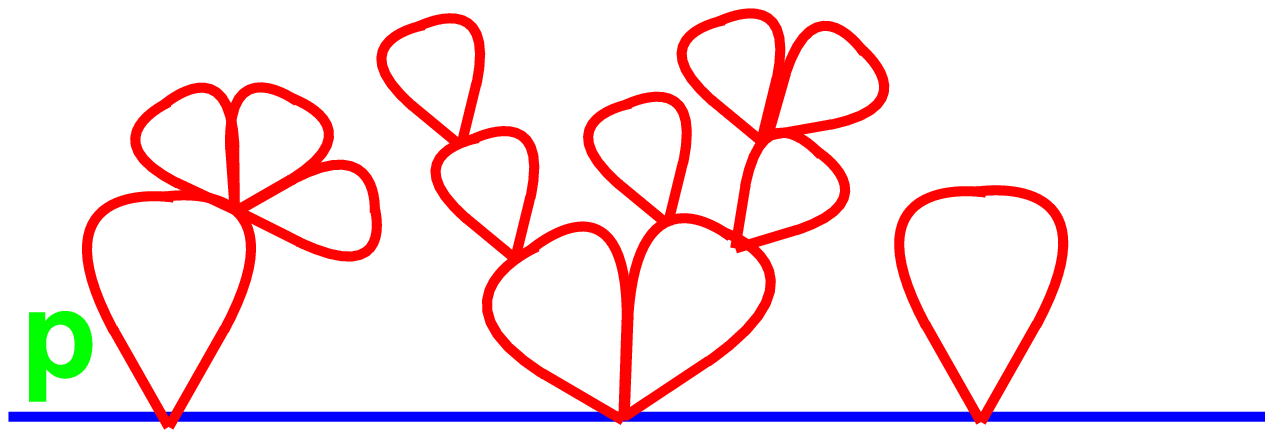}\includegraphics[width=\skpll,angle=-90,valign=c]{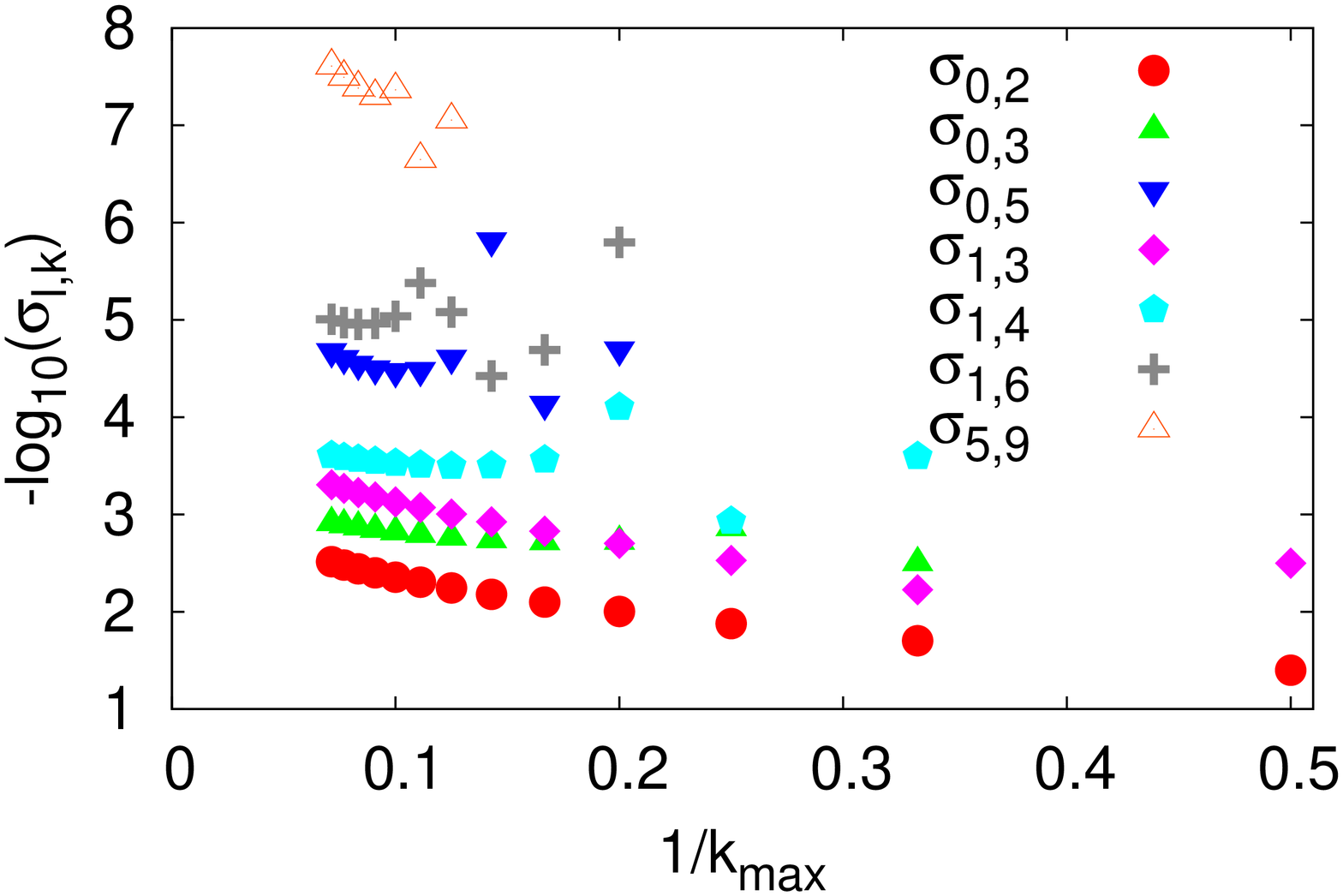}\\
  \includegraphics[width=\skpll,angle=-90]{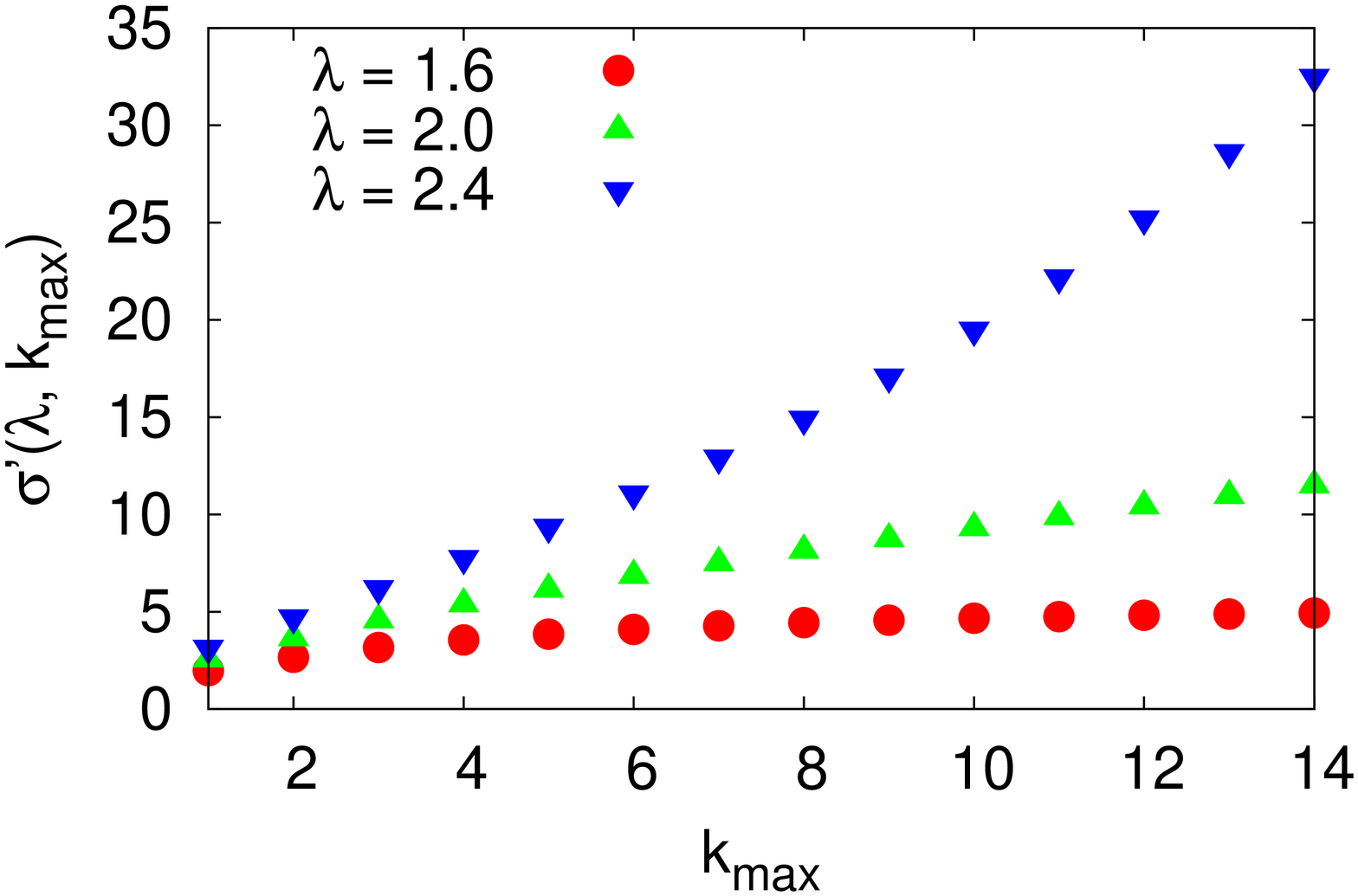}\includegraphics[width=\skpll,angle=-90]{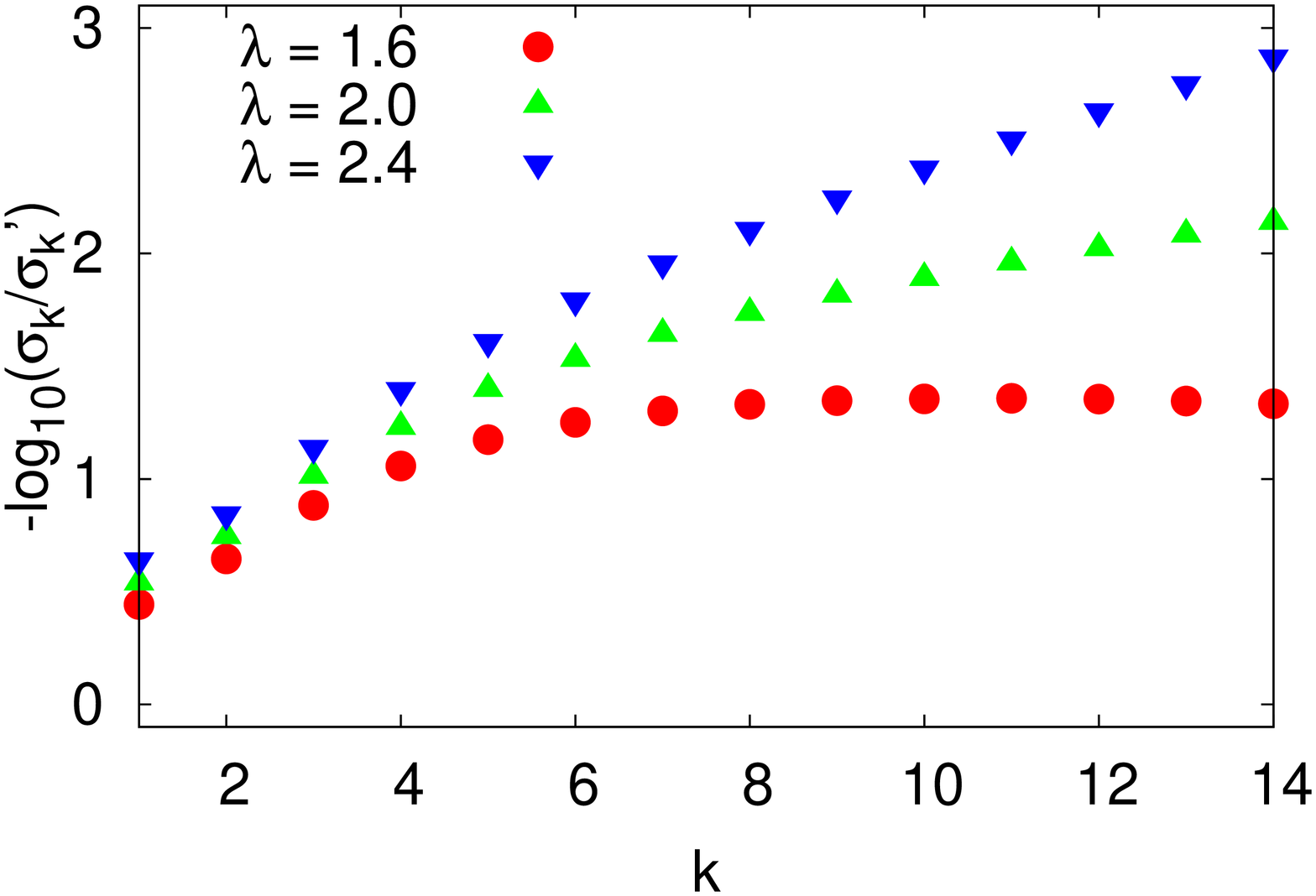}\\
  \caption{
  \textbf{Top left:} ``Cactus'' diagrams which contribute to the propagator $G\lr{p}$ in the Schwinger-Dyson equations (\protect\ref{perturbative_G_stereo_msummed}.)
  \textbf{Top right:} Dependence of the coefficients $\sigma_{k,l}$ in the double series (\protect\ref{xlogx_series}) on the truncation order $k_{max}$ in the formal power series (\protect\ref{formal_power_series}).
  \textbf{Bottom left:} result of summation over Feynman diagrams without taking into account the signs of their weights as a function of the truncation order $k_{max}$.
  \textbf{Bottom right:} effect of sign cancellations on the coefficients $\sigma_k$ of the formal power series (\protect\ref{formal_power_series}).}
  \label{fig:skl_convergence}
\end{figure*}
}{
\begin{figure*}[h!t]
  \centering
  \includegraphics[width=3.62cm,angle=-90]{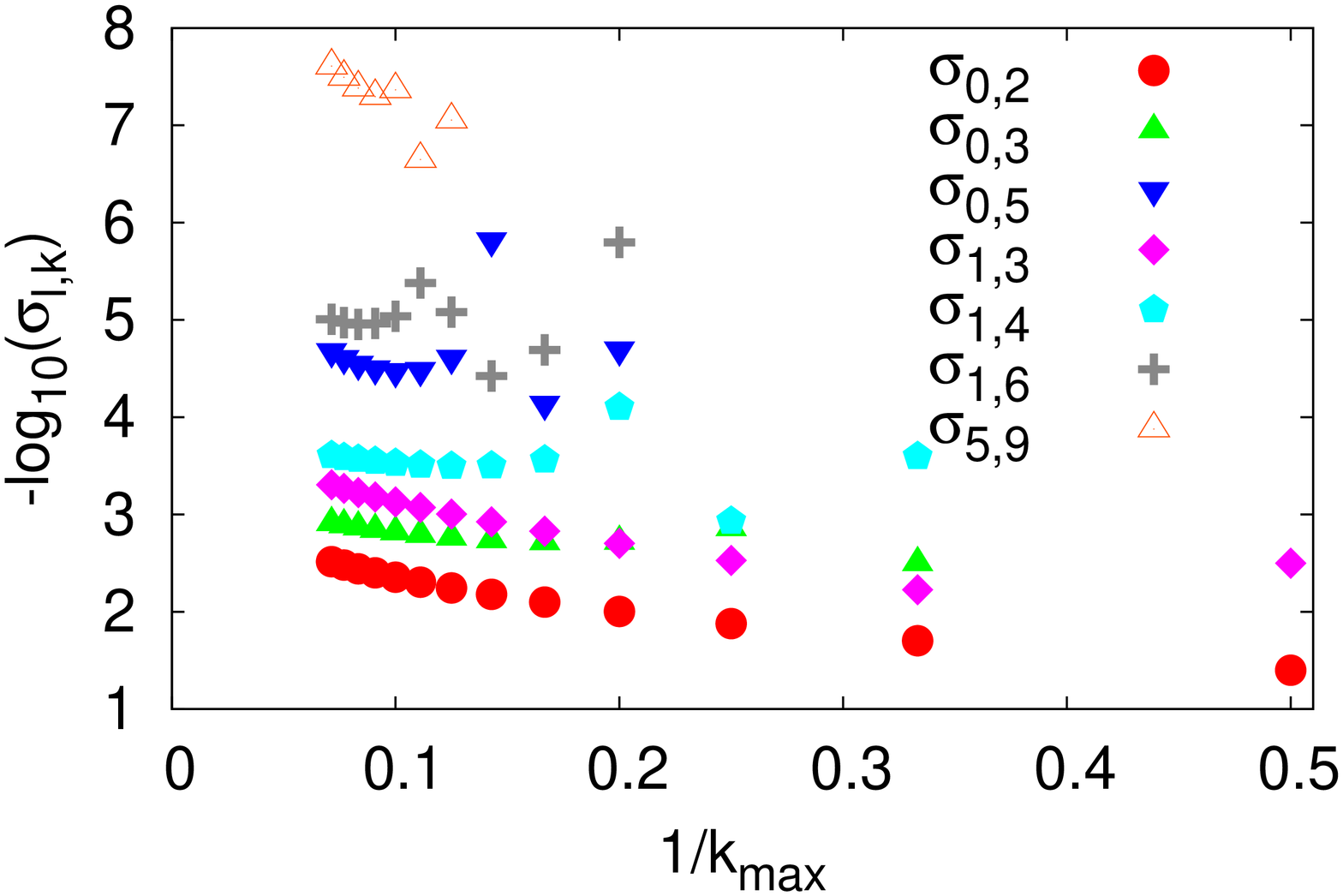}\hspace{-0.3cm}
  \includegraphics[width=3.62cm,angle=-90]{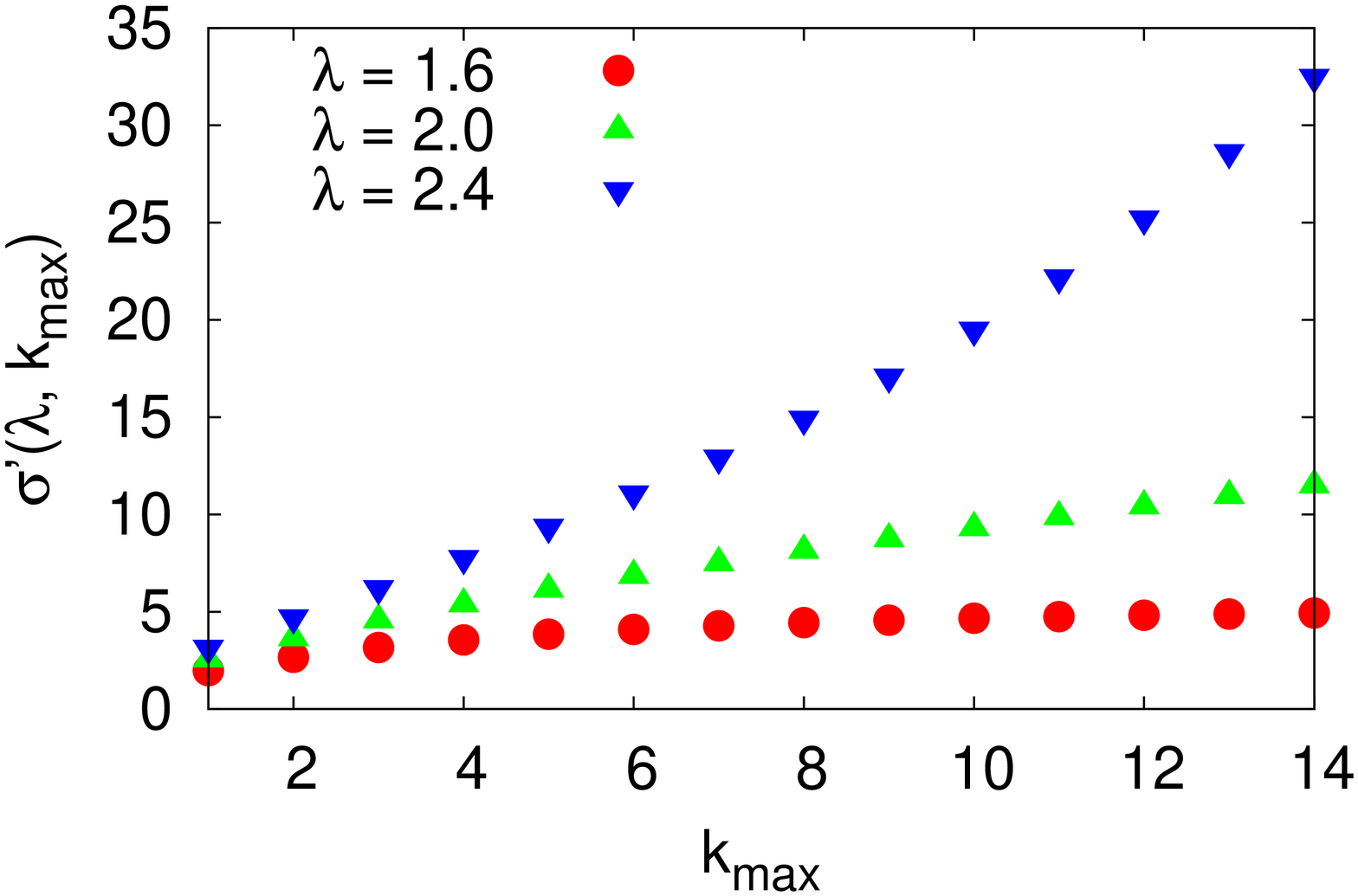}\hspace{-0.3cm}
  \includegraphics[width=3.62cm,angle=-90]{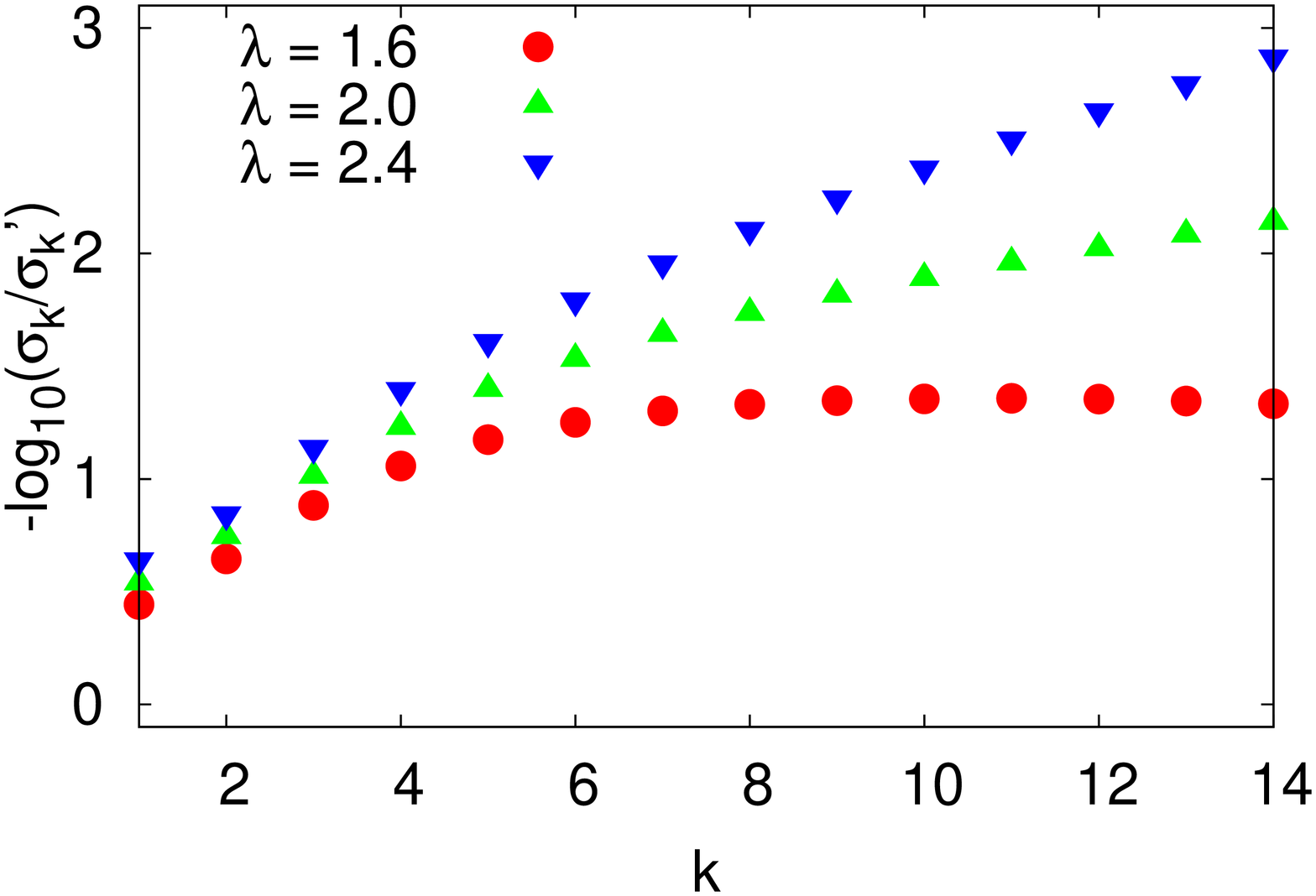}\\
  \caption{
   \textbf{On the left:} Dependence of the coefficients $\sigma_{k,l}$ in the double series (\protect\ref{xlogx_series}) on the truncation order $k_{max}$ in the formal power series (\protect\ref{formal_power_series}).
   \textbf{In the center:} result of summation over Feynman diagrams without taking into account their signs as a function of the truncation order $k_{max}$.
   \textbf{On the right:} effect of sign cancellations in the coefficients $\sigma_k$ of the formal power series (\protect\ref{formal_power_series}).}
  \label{fig:skl_convergence}
\end{figure*}
}

 The coefficients $z_k\lr{\sigma_0}$ and $\sigma_k\lr{\sigma_0}$ are now finite-degree polynomials in $\sigma_0$, $\sigma_0^{-1}$ and $\log\lr{\frac{\sigma_0}{32}}$ which can be represented in the following general form:
\begin{eqnarray}
\label{lambda_coeffs_sigma0}
 \sigma_k\lr{\sigma_0} =
 \sum\limits_{l=0}^{k}
 \lr{-\log\lr{\frac{\sigma_0}{32}}}^l
 \lr{\sum\limits_{i=i_{min}\lr{k-l-2}}^{\min\lr{l, 1}} c_{k,l,i} \sigma_0^i },
 \quad
 i_{min}\lr{j} =
 \left\{
   \begin{array}{ll}
       -j, & j \geq 0; \\
        0, & j  =   -1; \\
        1, & j  =   -2. \\
   \end{array}
 \right.  \quad ,
\end{eqnarray}
where each monomial term at fixed $k$, $l$, $i$ corresponds to a certain ``cactus''-like diagram.\ifarxiv{ Of course, if we have summed over a more general class of diagrams (e.g. planar diagrams), in the coefficients $\sigma_k\lr{\sigma_0}$ we would also encounter more complicated functions of $\sigma_0$ other than just logs and powers.}

 We obtain the coefficients $\sigma_k\lr{\sigma_0}$ up to some finite order $k_{max}$ using automated symbolic algebra and truncate the series (\ref{formal_power_series}) by summing over all orders $k$ up to $k_{max}$. After that, we substitute $\sigma_0 \rightarrow \lambda/2$. As a result, we obtain the truncated double series in powers of both $\lambda$ and $\log\lr{\lambda}$ of the form
\begin{eqnarray}
\label{xlogx_series}
 \sigma\lr{\lambda, k_{max}} =
 \sum\limits_{l = 0}^{k_{max}} \lr{-\log\lr{\frac{\lambda}{64}}}^l
 \sum\limits_{k = \min\lr{l+2,k_{max}}}^{k_{max} + \min\lr{l,1}} \sigma_{l,k} \lambda^k ,
\end{eqnarray}
and similarly for $z\lr{\lambda, k_{max}}$. It is important to stress that since upon the substitution $\sigma_0 \rightarrow \lambda/2$ the coefficients $\sigma_k\lr{\sigma_0}$ contain also negative powers of $\lambda$, even coefficients $\sigma_k\lr{\sigma_0}$ with large $k$ contribute to the coefficients $\sigma_{l,k}$ with small $k$ in the expansion (\ref{xlogx_series}), including the ones with $k = 0$. In order to check whether the coefficients $\sigma_{k,l}$ have some well-defined limit as $k_{max} \rightarrow \infty$, on the \iftoggle{arxiv}{top right}{leftmost} plot in Fig.~\ref{fig:skl_convergence} we show the dependence of some of these coefficients on $1/k_{max}$ at fixed $k$ and $l$. It seems that at least the few lowest-order coefficients do converge to certain limits at $k_{max} \rightarrow \infty$.

\ifarxiv{\subsection{Convergence of double series}
\label{subsec:double_series_convergence}}

 Next, we check how well do the series (\ref{formal_power_series}) truncated at the finite order $k_{max}$ approximate the exact results (\ref{on_exact_solution}). To this end on Fig.~\ref{fig:extrapolation_plots} we compare the results of the summation of truncated series for the renormalization factor $z\lr{\lambda, k_{max}}$ and the renormalized mass $m\lr{\lambda, k_{max}} = \sqrt{\sigma\lr{\lambda, k_{max}}}$ with exact results (\ref{on_exact_solution}). We see that the dynamically generated mass $m\lr{\lambda, k_{max}}$ converges quite quickly to the exact result, and linear extrapolation to $1/k_{max} = 0$ is enough to reproduce it within several percents.\ifarxiv{ It is also interesting that while at small $k_{max}$ the truncated sum is always larger than the exact result, at larger values of $k_{max}$ $m\lr{\lambda, k_{max}}$ develops some sort of bend, at which it goes below the exact result and then turns upward again. This bend is most clearly seen at large values of $\lambda$, however, it seems that at smaller values of $\lambda$ it still occurs at very small values of $1/k_{max}$, which we do not reach with our numerical recursion.

 In our opinion, such a quick convergence of the double series of the form (\ref{xlogx_series}) to the exact function $m\lr{\lambda}$ in (\ref{on_exact_solution}) which has an essential singularity at $\lambda = 0$ is a remarkable fact, and it would be interesting to understand the mathematical structure of the double series (\ref{xlogx_series}) and the possible relation to trans-series \cite{Unsal:14:2} in more details. In particular, let us notice that the perturbative expansion of the $O\lr{N}$ sigma model admits also the representation in terms of the general resurgent trans-series, which are the triple series in $\lambda$, $\log\lr{\lambda}$ and $e^{-S_0/\lambda}$, with $S_0$ being the action of unstable classical solutions \cite{Unsal:15:1}. In these triple series, sub-series in powers of $\lambda$ are divergent, and the whole expression only makes sense due to delicate cancellations between the ambiguities of Borel re-summation of sub-series in $\lambda$ and the factors $e^{-S_0/\lambda}$. However, if we perform the perturbative analysis using the representation (\ref{on_partition_perturbative}), there are additional terms in the action coming from the nontrivial integration measure (\ref{SN_integration_measure_stereo}) in terms of the fields $\phi_{i\,x}$. These terms are of course not present in the classical action, and hence the field configurations of kink and bion type \cite{Unsal:15:1} which are unstable solutions of the classical equations of motions are no longer the extrema of the action in (\ref{on_partition_perturbative}). It would be interesting to check whether the inclusion of the Jacobian into the action can completely eliminate non-perturbative saddle points. If this could be the case, it is not surprising that triple resurgent trans-series in powers of $\lambda$, $\log\lr{\lambda}$ and $e^{-S_0/\lambda}$ reduce to the double series of the form (\ref{xlogx_series}) which only involve $\lambda$ and $\log\lr{\lambda}$.

 A regularization of divergent power series somewhat similar to (\ref{xlogx_series}) was also considered in \cite{Prokofev:10:1}, where divergent series were approximated by another series involving fractional powers $\lambda^{1/m}$ of the coupling $\lambda$. In view of the identity $\log\lr{\lambda} = \lim\limits_{m \rightarrow \infty} m \lr{\lambda^{1/m} - 1}$, it is not unfeasible that the regularization of \cite{Prokofev:10:1} can be related to the double series (\ref{xlogx_series}). Note, however, that in contrast to \cite{Prokofev:10:1} in our case we do not need any additional extrapolation parameter which would control the convergence of the series - it is enough just to calculate the series to sufficiently high order.

} On the other hand, for the renormalization factor $z\lr{\lambda, k_{max}}$ the convergence to the exact answer $z\lr{\lambda} = 2$ is not so fast, and linear extrapolation from finite $k_{max} \sim O\lr{10}$ is not enough to reproduce it with sufficiently good precision. \ifarxiv{It seems that $z\lr{\lambda, k_{max}}$ has some weak logarithmic or fractional power singularity at $1/k_{max} = 0$.}

\begin{figure*}[!ht]
  \centering
  \newlength{\expll}\setlength{\expll}{5.29cm}
  \includegraphics[width=\expll,angle=-90]{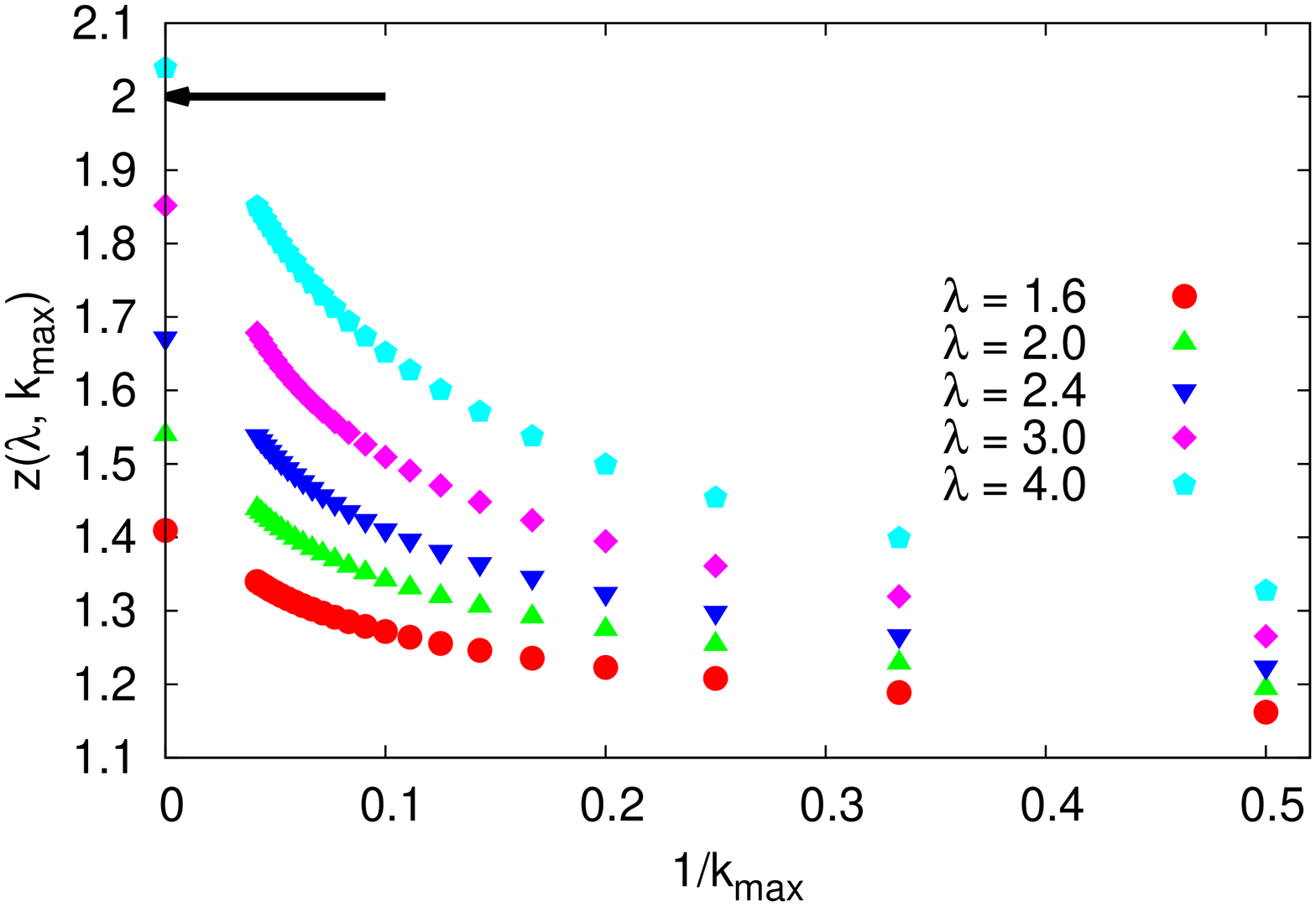}\includegraphics[width=\expll,angle=-90]{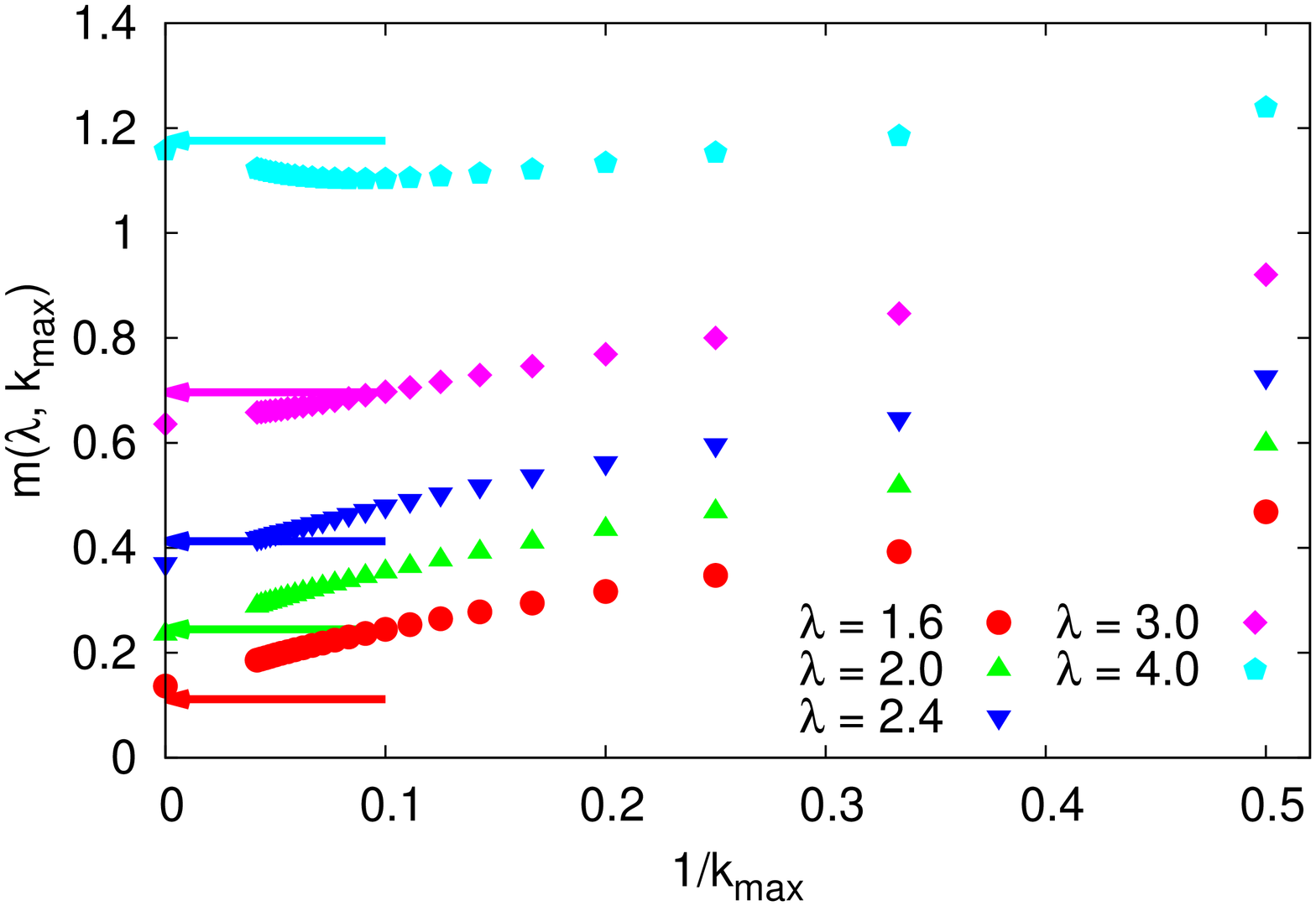}\\
  \ifarxiv{\includegraphics[width=\expll,angle=-90]{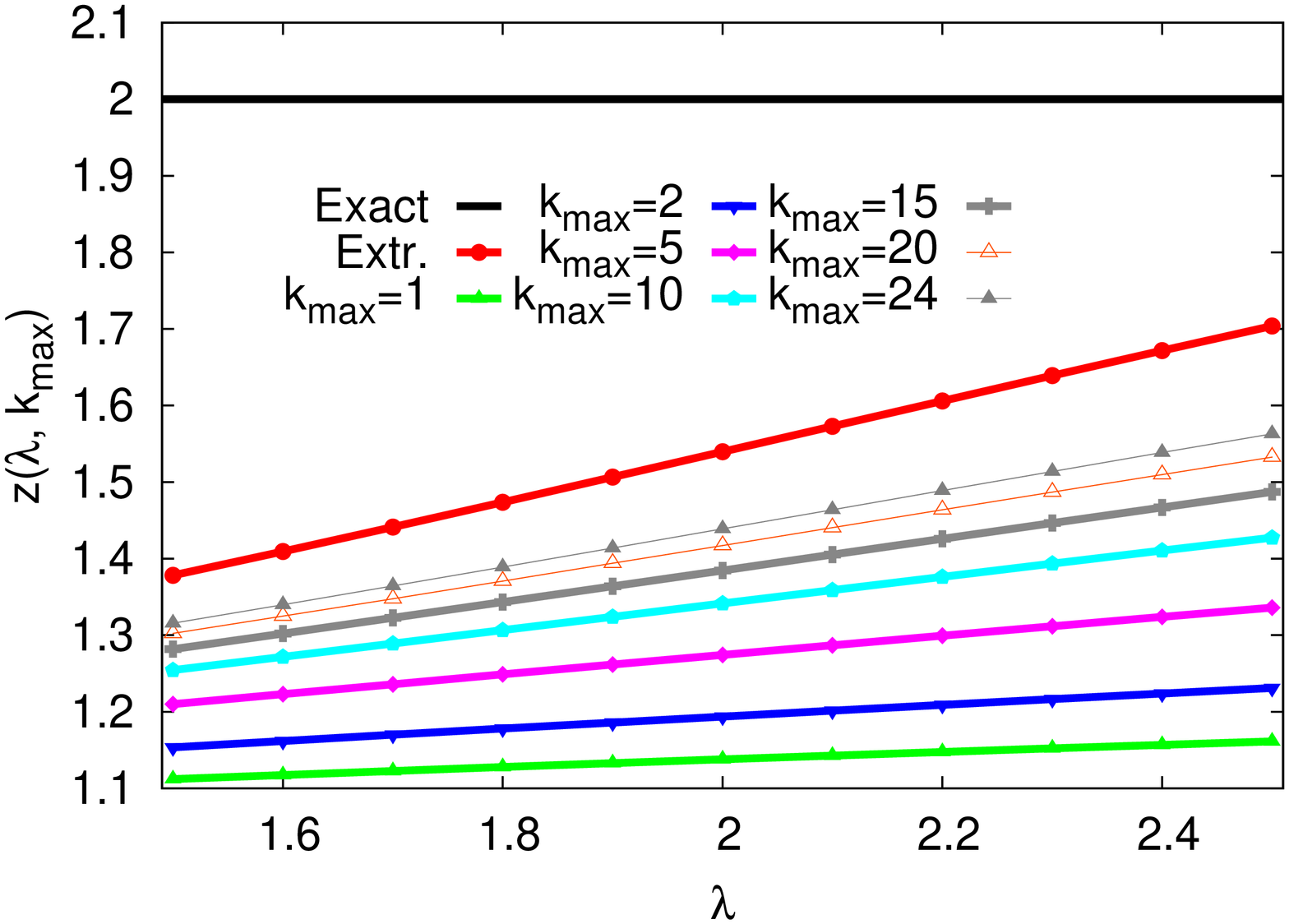}\includegraphics[width=\expll,angle=-90]{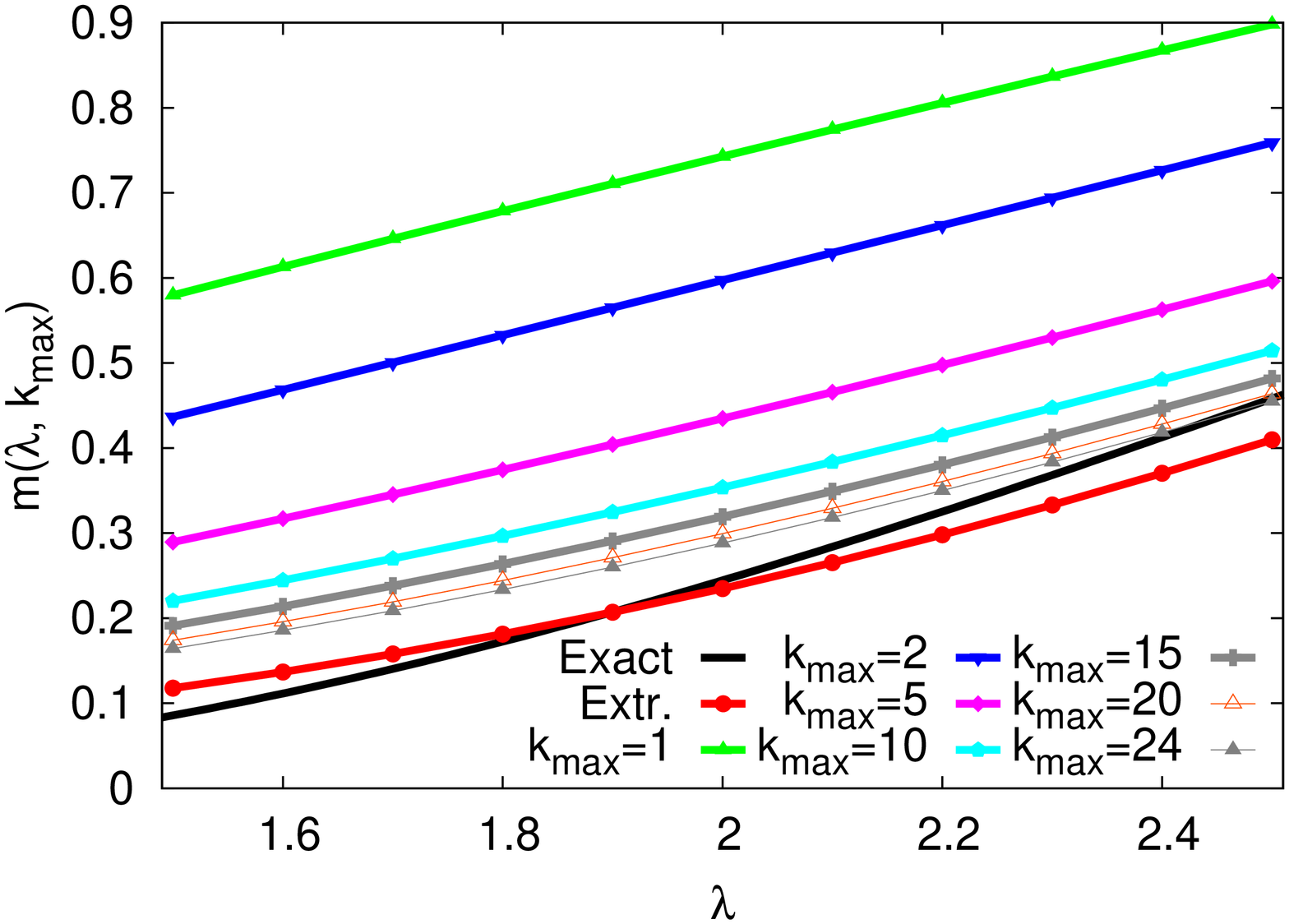}\\}
  \caption{\iftoggle{arxiv}{\textbf{At the top:} d}{D}ependence of the renormalization factor $z\lr{\lambda, k_{max}}$ (on the right) and the renormalized mass $m\lr{\lambda, k_{max}} \equiv \sqrt{\sigma\lr{\lambda, k_{max}}}$ (on the left) on the truncation order $k_{max}$ in the formal power series (\protect\ref{formal_power_series}). Points at $1/k_{max} = 0$ are the results of linear extrapolation from the numerical results with $7$ largest values of $k_{max}$. Arrows point to exact results (\protect\ref{on_exact_solution}). \ifarxiv{\textbf{At the bottom:} $z\lr{\lambda, k_{max}}$ (on the right) and $m\lr{\lambda, k_{max}}$ (on the left) as functions of $\lambda$ at different truncation orders $k_{max}$. The line labeled ``Extr.'' was obtained from the same extrapolation as for the plots at the top.}}
  \label{fig:extrapolation_plots}
\end{figure*}

\ifarxiv{\subsection{Sampling and reweighting in Diagrammatic Monte-Carlo}
\label{subsec:double_series_sampling}}

 Let us now imagine, that we are not aware of the exact solution (\ref{on_exact_solution}), and we try to find the coefficients $z_k\lr{\sigma_0}$ and $\sigma_k\lr{\sigma_0}$ in (\ref{formal_power_series}) by incorporating a generic DiagMC algorithm. Namely, we sample the bare Feynman diagrams with the probability proportional to their weight. For our model, the weights of individual ``cactus''-like Feynman diagrams \ifarxiv{(with internal momenta already integrated out)} correspond to the coefficients of the monomial terms in (\ref{lambda_coeffs_sigma0}). Since in our case these coefficient can be both positive and negative, for Monte-Carlo sampling we could take the absolute value of the diagram weight and treat the sign by reweighting. We denote the sum in (\ref{lambda_coeffs_sigma0}) with all coefficients $c_{k,l,i}$ taken by absolute value as $\sigma_k'\lr{\sigma_0}$. Inserting the coefficients $\sigma_k'\lr{\sigma_0}$ into the series (\ref{formal_power_series}) instead of $\sigma_k\lr{\sigma_0}$ and restricting the summation over $k$ to $k \leq k_{max}$, we obtain the function $\sigma'\lr{\lambda, k_{max}}$.

 We first check whether the sum of the absolute values of diagram weights is finite in the limit $k_{max} \rightarrow \infty$, so that individual weights can be interpreted as probabilities. To this end, on the \iftoggle{arxiv}{bottom left}{central} plot on Fig.~\ref{fig:skl_convergence} we plot $\sigma'\lr{\lambda, k_{max}}$ as a function of $k_{max}$. This plot indicates that $\sigma'\lr{\lambda, k_{max}}$ has no finite limit at $k_{max} \rightarrow \infty$, and the resulting double series are divergent unless one takes into account sign cancellations. In practice, however, this divergence can be quite easily circumvented by separately sampling the diagrams of different order and subsequent explicit summation\ifarxiv{ (see e.g. \cite{Buividovich:11:1})}.

 \iftoggle{arxiv}{Provided one can deal with the divergence of the series being sampled, t}{T}he next question is then how strong are cancellations between same-order diagrams with positive and negative weights. To answer this question, on the \iftoggle{arxiv}{right bottom}{rightmost} plot on Fig.~\ref{fig:skl_convergence} we show the ratios of the coefficients $\sigma_k\lr{\sigma_0}$ (where diagram weights retain their sign) to $\sigma_k'\lr{\sigma_0}$ (where the diagram weights are summed by absolute value). If this ratio is small, then sign cancellations are important. We see that the sign cancellations are quite important for high orders $k$, where the absolute values of $\sigma_k\lr{\sigma_0}$ and $\sigma_k'\lr{\sigma_0}$ differ two or three orders of magnitude. However, at small values of $\lambda$ the sign cancellations become milder and only weakly depend on the diagram order $k$. Since small values of $\lambda$ correspond to the physically most interesting continuum limit, this is a very promising observation. \ifarxiv{Of course, one should still keep in mind that according to Fig.~\ref{fig:extrapolation_plots} one should take into account more and more terms in the expansion when approaching the continuum limit, but this is not a principal problem for DiagMC algorithms.}

\section{Conclusions}
\label{sec:conclusions}

 In these Proceedings, we have considered the weak-coupling perturbative expansion of the $O\lr{N}$ sigma-model in the large-$N$ limit, taking the practical perspective of sampling Feynman diagrams by a generic DiagMC algorithm. In order to set up the perturbative expansion, we have used the stereographic mapping. As a result, bare propagators have acquired a small bare mass term proportional to the coupling $\lambda$. Counting only the positive powers of $\lambda$ associated with interaction vertices, we have arrived at the double series representation which involves powers of both $\lambda$ and $\log\lr{\lambda}$. We have numerically checked the convergence of these series to the exact results, which turned out to be particularly fast for the dynamically generated mass gap and slower for the renormalization factor. Moreover, we have demonstrated that it is feasible to obtain the relevant series coefficients by a fictitious Monte-Carlo sampling in the space of bare Feynman diagrams. Interestingly, the sign problem which appears in such a sampling becomes milder as we approach the continuum limit at $\lambda \rightarrow 0$.
\ifarxiv{}
 These observations are very promising for further applications of DiagMC algorithms to asymptotically free field theories, most notably for the $U\lr{N}$ principal chiral models and for non-Abelian gauge theories on the lattice. \ifarxiv{Similarly to the stereographic mapping (\ref{stereographic_mapping_perturbative}) from the sphere $S_N$ to the real space $\mathbb{R}^{N-1}$, one can also use the stereographic mapping $g = \frac{1 - i \phi}{1 + 1 \phi}$ from $U\lr{N}$ group to the (whole) space of Hermitian matrices $\phi$, which would also result in a bare mass term proportional to $\lambda$ once the nontrivial integration measure is included into the action \cite{Buividovich:10:5}. Let us note that the exponential mapping $g = e^{i \phi}$, which is most commonly used for constructing lattice perturbation theory \cite{DiRenzo:04:1}, also results in a small addition to the quadratic part of the action. However, this additional term has the double-trace structure and one flat direction in the space of $\phi$, which makes the automated construction of perturbation theory more complicated (for an explicit demonstration, see Appendix \ref{apdx:un_exp_mapping_measure}).}

%\bibliographystyle{mybibstyle}
%\bibliography{Buividovich}
\iftoggle{arxiv}{

}{

}

\iftoggle{arxiv}{
\appendix

\section{Integration measure on the sphere $S_N$ in terms of stereographic coordinates}
\label{apdx:sn_stereo_mapping_measure}

 In order to find the integration measure on the $N$-sphere $S_N$ in terms of the coordinates $\phi_i$ introduced in (\ref{stereographic_mapping_perturbative}), we first find the metric form on $S_N$ in terms of $\phi_i$:
\begin{eqnarray}
\label{sn_stereo_metric}
 ds^2 = d n_0^2 + d n_i^2
 =
 \frac{\lambda^2 \lr{\phi_i d\phi_i}^2}{\lr{1 + \frac{\lambda}{4} \phi^2}^4}
 +
 \lr{
 \frac{\sqrt{\lambda} d\phi_i}{1 + \frac{\lambda}{4} \phi^2}
 -
 \frac{\lambda/2 \phi_j d\phi_j \sqrt{\lambda} \phi_i}{\lr{1 + \frac{\lambda}{4} \phi^2}^2}
 }^2
 = %\nonumber \\ =
 \frac{\lambda d\phi_i^2}{\lr{1 + \frac{\lambda}{4} \phi^2}^2} .
\end{eqnarray}
We see that the metric tensor $g_{ij}$ in $ds^2 = g_{ij} d\phi_i d\phi_j$ appears to be diagonal in the stereographic coordinates $\phi_i$ (which actually stems from the fact that stereographic mapping is a conformal one), and thus we can immediately find the integration measure:
\begin{eqnarray}
\label{sn_stereo_measure}
 \int\limits_{S_N} dn
 =
 \int\limits_{\mathbb{R}^{N-1}} d^{N-1}\phi \, \sqrt{\det{g}}
 =
 \int\limits_{\mathbb{R}^{N-1}} d^{N-1}\phi \, \lr{1 + \frac{\lambda}{4} \phi^2}^{-\lr{N-1}}.
\end{eqnarray}
Note that in contrast to the exponential mapping discussed in Appendix \ref{apdx:un_exp_mapping_measure}, on the r.h.s. of the above expression the integration should be performed over the whole $N-1$ dimensional real space $\mathbb{R}^{N-1}$. If one is interested only in the leading order of the expansion in $1/N$, one can also replace the power of $-\lr{N-1}$ in (\ref{sn_stereo_measure}) by $-N$, as in (\ref{SN_integration_measure_stereo}).

\section{Schwinger-Dyson equations for $O\lr{N}$ sigma model in stereographic coordinates}
\label{apdx:sd_equations}

 Schwinger-Dyson equations for the two-point function $\vev{\phi_x \phi_y}$ in the $O\lr{N}$ sigma-model can be derived by expanding the following full derivative in the path integral:
\begin{eqnarray}
\label{SD_equations_start}
 \mathcal{Z}^{-1} = \int \mathcal{D}\phi
 \frac{\partial}{\partial \phi_{i \, x}} \lr{
  \phi_{i \, y} \, e^{-S\lrs{\phi}}
 } = 0 .
\end{eqnarray}
We now use the explicit form of the action from (\ref{on_partition_perturbative}) and expand the full derivative. We also use the factorization property of $O\lr{N}$-singlet observables $\vev{\lr{\phi_{x} \cdot \phi_{y}} \lr{\phi_{z} \cdot \phi_{t}}} = \vev{\phi_{x} \cdot \phi_{y}} \vev{\phi_{z} \cdot \phi_{y}} + O\lr{1/N}$ as well as translational invariance of all the observables which implies that $\vev{\phi_{x} \cdot \phi_{y}}$ depends only on the difference $x - y$. As a result, we obtain
\begin{eqnarray}
\label{SD_equations_xspace}
 \sum\limits_z \lr{D_{x z} + \frac{\lambda}{2} \delta_{x,z}} \, \vev{\phi_z \cdot \phi_y}
 + \nonumber \\ +
 \sum\limits_{\substack{k,l=0\\k+l\neq 0}}^{+\infty}
   \frac{\lr{-1}^{k+l} \lambda^{k+l}}{4^{k+l}}
   \vev{\phi_x \cdot \phi_y} \sum\limits_z D_{x z} \vev{\phi_z \cdot \phi_x}
   \lr{k + l} \vev{\phi_x^2}^{k+l-1}
  + \nonumber \\ +
  \sum\limits_{\substack{k,l=0\\k+l\neq 0}}^{+\infty}
   \frac{\lr{-1}^{k+l} \lambda^{k+l}}{4^{k+l}}
   \vev{\phi_x^2}^{k+l} \sum\limits_z D_{x z} \vev{\phi_z \cdot \phi_y}
 +
 2 \sum\limits_{k=2}^{+\infty} \frac{\lr{-1}^{k-1} \, \lambda^k}{4^k} \vev{\phi_x^2}^{k-1} \vev{\phi_x \cdot \phi_y} = 1 .
\end{eqnarray}
We note that all the summands in the double sums over $k$ and $l$ depend only on the sum $k+l$, therefore they can be reduced to single sums over a single integer $m = k+l$. It is also convenient now to use the momentum-space representation $\vev{\phi_x \cdot \phi_y} = \int \frac{d^2 p}{\lr{2 \pi}^2} e^{i p \lr{x - y}} G\lr{p}$ of the two-point function. We find, eventually, the following equations
\begin{eqnarray}
\label{perturbative_SD_stereo}
 G^{-1}\lr{p} = D\lr{p} + \frac{\lambda}{2}
 +
 D\lr{p} \, \sum\limits_{m=1}^{+\infty} \frac{\lr{-1}^m \lr{m+1}}{4^m} \, \lambda^m \, \lr{G_{zz}}^m
 + \nonumber \\ +
 \sum\limits_{m=2}^{+\infty} \frac{2 \lr{-1}^{m-1}}{4^m} \, \lambda^m \, \lr{G_{zz}}^{m-1}
 +
 \sum\limits_{m=1}^{+\infty} \frac{\lr{-1}^m \, m \lr{m+1}}{4^m} \, \lambda^m \, \lr{G_{zz}}^{m-1} \lr{D G}_{zz} ,
\end{eqnarray}
where $G_{zz} = \int\frac{d^2 p}{\lr{2 \pi}^2} G\lr{p}$ and $\lr{D G}_{zz} = \int\frac{d^2 p}{\lr{2 \pi}^2} D\lr{p} \, G\lr{p}$. Performing now the sums over $m$ explicitly, we arrive at the equations (\ref{perturbative_G_stereo_msummed}) in the main text of these Proceedings.

\section{$U\lr{N}$ Haar measure with exponential mapping}
\label{apdx:un_exp_mapping_measure}

 We consider the exponential mapping $g = e^{i \phi}$ from the space of Hermitian matrices $\phi$ to the space of unitary matrices $g \in U\lr{N}$. In order to express the group-invariant integration measure on $SU\lr{N}$ in terms of $\phi$, let us first find the metric form $d s^2 = \tr\lr{d g \, d g^{\dag}}$ in terms of $\phi$. To this end we use the identity $d e^{i \phi} = \int\limits_{0}^{1} dz \, e^{i z \phi} d\phi e^{i \lr{1 - z} \phi}$ to arrive at
\begin{eqnarray}
\label{un_metric1}
 \tr\lr{d g \, d g^{\dag}} = \int\limits_{0}^{1} \int\limits_{0}^{1} dz_1 dz_2 \,
 \tr\lr{e^{i \lr{z_2 - z_1} \phi} \, d\phi \, e^{-i \lr{z_2 - z_1} \phi} d\phi}  .
\end{eqnarray}
We now use the fact that the $U\lr{N}$ Haar measure which we are about to calculate is invariant under the similarity transformations $\phi \rightarrow u \phi u^{\dag}$, $u \in U\lr{N}$ and hence the measure depends only on the eigenvalues $\lambda_i$ of $\phi$. When calculating the determinant of the metric (\ref{un_metric1}) we can therefore assume that $\phi = \diag{\lambda_1, \ldots, \lambda_N}$. Then the metric form can be written as
\begin{eqnarray}
\label{un_metric2}
 \tr\lr{d g \, d g^{\dag}} =
 \sum\limits_{i,j}
 \int\limits_{0}^{1} \int\limits_{0}^{1} dz_1 dz_2 \,
 e^{i \lr{z_2 - z_1} \lr{\lambda_i - \lambda_j}} |d \phi_{i j}|^2
 = \nonumber \\ =
 \sum\limits_{i} d\phi_{ii}^2
 +
 2 \sum\limits_{i > j}
 \int\limits_{0}^{1} \int\limits_{0}^{1} dz_1 dz_2 \,
 \cos\lr{\lr{z_2 - z_1} \lr{\lambda_i - \lambda_j}} |d \phi_{i j}|^2
 = \nonumber \\ =
 \sum\limits_{i} d\phi_{ii}^2
 +
 8 \sum\limits_{i > j}
 \frac{\sin^2\lr{\lr{\lambda_i - \lambda_j}/2}}{\lr{\lambda_i - \lambda_j}^2} |d \phi_{i j}|^2,
\end{eqnarray}
where the indices $i$, $j$ run from $1$ to $N$ and in the last line we have already performed the integrations over $z_1$ and $z_2$. We thus see that upon the diagonalization of $\phi$ the metric form also becomes diagonal. Hence the invariant integration measure, which is the square root of the determinant of the metric form, is the product of the diagonal elements (and we have to remember that $d \phi_{ij}$ for $i > j$ is a complex number and hence has two components). Thus we arrive at the following general expression for the $SU\lr{N}$ Haar measure in terms of Hermitian matrices $\phi$:
\begin{eqnarray}
\label{SUN_haar_measure_phi}
 \int\limits_{U\lr{N}} dg
 =
 \const \, \int\limits_{\mathbb{M}} d\phi
 \prod\limits_{i > j}
 \frac{\sin^2\lr{\lr{\lambda_i - \lambda_j}/2}}{\lr{\lambda_i - \lambda_j}^2} ,
\end{eqnarray}
where $\const$ denotes an overall $\phi$-independent normalization factor and $\mathbb{M}$ is some $N^2$-dimensional region in the $N^2$-dimensional space $\mathbb{R}^{N^2}$ of the Hermitian $N \times N$ matrices $\phi$. The $U\lr{N}$ group manifold is uniquely covered by the integration in (\ref{SUN_haar_measure_phi}) if $\mathbb{M}$ is bounded by the $N^2-1$-dimensional manifolds at which at least two eigenvalues of $\phi$ coincide. E.g. in the case of $U\lr{2}$ group, $\mathbb{M}$ belongs to the four-dimensional real space $\mathbb{R}^4$ and is the direct product of the $3$-dimensional ball and the one-dimensional circle $S_1$.

 It is also instructive to reduce the integration over $\phi$ in the above expression to integration over eigenvalues $\lambda_i$ (which is reasonable for $U\lr{N}$-invariant functions which depend only on the eigenvalues of $\phi$). We then arrive at the expectable result
\begin{eqnarray}
\label{SUN_haar_measure_phievals}
 \int\limits_{U\lr{N}} dg
 =
 \const \, \int\limits_{\mathbb{M}} \prod\limits_i d\lambda_i
 \prod\limits_{i > j}
 \sin^2\lr{\lr{\lambda_i - \lambda_j}/2} .
\end{eqnarray}

 We can now try to incorporate the measure (\ref{SUN_haar_measure_phi}) into the action, writing it as
\begin{eqnarray}
\label{SUN_haar_measure_phi_exponentiated}
\int\limits_{\mathbb{M}} d\phi
 \expa{\sum\limits_{i, j}
 \log\lr{\frac{\sin^2\lr{\lr{\lambda_i - \lambda_j}/2}}{\lr{\lambda_i - \lambda_j}^2}}
 }
 =
 \int\limits_{\mathbb{M}} d\phi \expa{- \sum\limits_{i, j} \frac{\lr{\lambda_i - \lambda_j}^2}{12} + O\lr{\lambda^4}  }
 = \nonumber \\ =
\int\limits_{\mathbb{M}} d\phi \expa{- \frac{1}{6} \lr{N \tr\phi^2 - \tr\phi \, \tr\phi} + O\lr{\phi^4}  } .
\end{eqnarray}
We thus see that for the exponential map $g = e^{i \phi}$ the exponentiated measure also includes the terms which are quadratic in $\phi$ and which will therefore produce some bare mass terms in the bare propagators of the field $\phi$. It is also interesting to see the appearance of the double-trace term $\tr\phi \, \tr\phi$. However, this quadratic part of the integration measure has a flat direction for which $\phi$ is proportional to the identity matrix, which might still lead to IR divergences. Furthermore, when constructing perturbative expansions, one typically extends the integration from $\mathbb{M}$ to the whole real space $\mathbb{R}^{N^2}$ of Hermitian matrices. We see that with the exponential map such extension results in multiple covering of the $U\lr{N}$ group manifold, which should be amended by introducing ghost fields. It seems thus that the exponential mapping would lead to a more complicated structure of the perturbative expansion than the stereographic one (see \cite{Buividovich:10:5} for the derivation of the $SU\lr{N}$ integration measure in terms of stereographic coordinates).
}{}

\end{document}